# Mode volume of electromagnetic resonators: let us try giving credit where it is due

Philippe Lalanne

Laboratoire Photonique, Numérique et Nanosciences (LP2N), IOGS-University of Bordeaux-CNRS, 33400 Talence cedex, France

**Abstract**: The mode volume is a very important figure of merits, which transversally arouse the physics and applications of electromagnetic cavities. Its modern definition was not obtained simply, and I try to briefly summarize the evolution of ideas. I also incidentally address considerations on scientific publishing policy in the epilogue.

## 1. Introduction

Two characteristic parameters, which figure prominently in the physics and device applications of cavities, quantify the ability of cavity modes (we will call them quasinormal modes or QNMs hereafter to emphasize their non-Hermitian character due to leakage) to boost light-matter interactions, the quality factor $Q$ and the mode volume $V$ [1,2]. $Q$ governs the enhancement due to temporal confinement. $V$ is an important notion, at least as important, which determines the enhancement due to spatial confinement and is directly related to the mode normalization, which prefigures how much a mode is excited by an external source.

To figure out the significance of $V$, imagine any microcavity perturbed by a tiny foreign object, a point-like molecule for example. The interaction is bidirectional: on one side, the molecule perturbs the cavity QNMs, therein modifying their complex frequencies, and on another side, the modes polarize the molecule. The strength of this two-way interaction, which depends on the position $r$ of the molecule, measures the mode capability to strengthen light-matter interaction; it is the mode volume.

In a recent Tutorial published in Adv. Opt. Photon. [3], the mode volume notion is introduced by the following statement (the quote is not taken out of context, see Section 1.2 in [3]):

> *«… the result of a QNM model of the Purcell factor can be immediately written in exactly the same form of the original formula due to Purcell [[16]], but with a slightly modified expression for the effective mode volume [[17]]; this definition of a mode volume for a leaky resonator can even be extended to plasmonic systems by properly accounting for absorption and dispersion in the material [[18]].»* where [[16]] is the famous half-page 1946 report by E. M. Purcell, [[17]] is a 2012 Optics Letters manuscript referred to as [4], and Ref. [[18]] is a 2013 Physical Review Letters referred to as [5] hereafter.

The goal of this short report is to examine the exactness of the 'extension statement' presenting the work performed in [5] as an extension of [4]. Potential readers may find me picky on the use of words. However, I believe that this is not the case on that occasion. The exactness of the statement has been debated by Philip Kristensen at the end of two talks I gave in June 2019 at PIERS Rome [6], and more recently in an online symposium [7]. The latest vivid debate shows that a consensus is not reached. And the debate started long ago. In January 2014, I sent an email to Harry Atwater, chief editor of ACS Photonics, with a detailed commentary of a perspective article posted online by ACS Photonics and authored by Philip Kristensen and his colleagues [8]. My comments have not been taken lightly by the editors: the posted initial version has been quite revised before final publication [9] to remove some of my concerns and six months later in June, when *ACS Photonics* was featuring

Comment manuscript type, Harry Atwater invited me to submit a Comment. We did not take the opportunity.[1]

Beyond any personal divergence opinions that are considered in the epilogue, it is always instructive to clarify how the evolution of ideas occurs in the process of science. I expect that this report contributes to this type of clarification.

## 2. Mode volume: a brief historical perspective

QNMs have been computed for many decennia, with in-house or commercial electromagnetic software. However, it is only since a few years that they offer normalized modes. The reason is that QNMs are far more complicated than normal modes of Hermitian bound systems, simply because QNM fields of open non-Hermitian systems satisfy the outgoing-wave conditions and exponentially grow far away from the resonators. This is very simple to see by noting that fields far away from resonators typically behave as $\exp[-i\widetilde{\omega}(t - r/c)]/r$, where $\widetilde{\omega}$ is the QNM complex frequency. Since they exponentially decay in time as $t \to \infty$, QNMs exponentially grow in space as the distance $r$ from the resonator in free space tends towards infinity, see Section 2.1 in [2]. Therefore, QNMs are exponentially divergent wavefunctions which by definition do not belong to the Hilbert space of conventional Hermitian systems taught in quantum mechanics books for instance. Therefore, classical inner products (orthogonality or normalization) generally applicable to normal modes do not apply. This is unfortunate as normalized field distributions allow a direct evaluation of the force with which QNM fields interact with matter.

**Hermitian V's of conservative systems.** The concept of mode volume was initially introduced for high-$Q$ systems by pioneering works on the perturbation theory of radiofrequency cavities [10] and the control of spontaneous emission at microwave frequencies [11]. It is thus quite naturally that the community opted for a definition of the mode volume well-established by Hermitian theory of conservative systems,

$$V(\boldsymbol{r}) = \frac{\int \left[\varepsilon|\widetilde{\mathbf{E}}|^2 + \mu_0|\widetilde{\mathbf{H}}|^2\right] d^3 r}{2\varepsilon(r)|\widetilde{\mathbf{E}}(r)\cdot\mathbf{u}|^2}, \tag{1}$$

where **u** is the polarization direction of the tiny perturber (or quantum emitter) and $V$ is a real-valued positive quantity that suggests that the strongest interaction is implemented at the field-intensity maximum. A didactic presentation of the Hermitian theory of normal modes in electromagnetism can be found in the chapter 2 in [12] for instance.

The Hermitian definition has been successfully used for quantum electrodynamics studies of various solid-state microcavities, e.g. micropillars, photonic-crystal cavities [1,13] or related studies on cavity perturbation theory [2] or lasers [14]. Indeed, for an exponentially growing field, the Hermitian volume integral is diverging as the domain of integration of the integral in the numerator expands. In practice, the integration domain is thus truncated arbitrarily, for instance with Born-von Kármán periodic boundary conditions [13]. Although not intellectually satisfactory, this solution is approximately safe for high-$Q$ photonic-crystal cavities, because the leakage is weak and the exponential growth is only effective at large distances $\sim Q\lambda$ from the cavity. This explains why, in practice for high-$Q$ modes in dielectric constructs, $V$ is often computed with the energy-based definition, and this has been done extensively over the past 20 years for photonic-crystal cavities.

**Non-Hermitian V's of non-conservative systems.** The Hermitian $V$ is accurate to predict the coupling strength with a high-$Q$ QNM, but it is incapable to predict any non-Hermitian effects. For instance, first-order perturbation theory of normal modes cannot predict the $Q$-change of a photonic-crystal cavity due to a tiny perturbation; it can just predict the frequency shift [15,2]. For plasmonic nanocavities, the Hermitian $V$ is completely unusable.

---

[1] By referring to this old story, my aim is primarily to fully present the context. The longstanding absence of evolution in the presentation of V's ideas also explains why the style of the present report, albeit factual, is a bit direct in certain occasions. The annotated pdf version of the Just Accepted Manuscript [8] which I sent to H. Atwater, S. Hughes and P. Kristensen can be provided on request. The notes instructively highlight how much our 2013 works [5] and [27] have been obscured in the initial online version.

Nowadays, the mode volume is defined as a complex-valued quantity [2] *symbolically* written

$$\tilde{V}(\mathbf{r}) = \frac{\int \left[\tilde{\mathbf{E}}_m \cdot \frac{\partial \omega \varepsilon}{\partial \omega} \tilde{\mathbf{E}}_m - \tilde{\mathbf{H}}_m \cdot \frac{\partial \omega \mu}{\partial \omega} \tilde{\mathbf{H}}_m \right] d^3 r}{2\varepsilon(\mathbf{r})[\tilde{\mathbf{E}}(\mathbf{r}) \cdot \mathbf{u}]^2} , \qquad (2a)$$

on the condition *that a hidden difficulty*, the square integrable numerator,

$$\int \left[\tilde{\mathbf{E}}_m \cdot \frac{\partial \omega \varepsilon}{\partial \omega} \tilde{\mathbf{E}}_m - \tilde{\mathbf{H}}_m \cdot \frac{\partial \omega \mu}{\partial \omega} \tilde{\mathbf{H}}_m \right] d^3 r \text{ is defined properly.} \qquad (2b)$$

Without Eq. (2b), Eq. (2a) is meaningless due to divergence of the integrand volume. Note that, compared to Eq. (1), the integrand is written to account for anisotropic and dispersive materials.

A QNM is a source-free solution of Maxwell's equations, and thus if $\tilde{\mathbf{E}}$ is a solution $\gamma \tilde{\mathbf{E}}$ for a complex valued $\gamma$ is also a solution. Knowing how much a tiny perturber acts on a QNM field requires a normalization. It is the role of the numerator integral in Eq. (2a). Over time, the literature has presented a series of normalization methods. Some of them are mathematically correct, some others are correct under certain restrictions, some other are almost always false. This is the difficult normalization issue discussed in the next section.

In the early 10's, two letters have dominantly contributed to raise the issue of mode volumes for non-Hermitian electromagnetic resonators. The first contribution is a letter published in 2012 by Philip Kristensen and his colleagues [4]. The letter first recalls the classical Hermitian definition of $V$, Eq. (1), discusses its inconsistency for non-Hermitian systems and then proposes to adopt a normalization proposed in an earlier work [16] of the Hong Kong group of P. T. Leung, K. Young and colleagues. The works of this group will be further discussed in the next Section. The normalization encompasses a surface integral and a volume integral, see Eq. (4) in [4]. From this, directly follows the prescription of a complex mode volume, Eq. (5) in [4]. The letter is easy to read. It focuses on the real part of $\tilde{V}$ and shows mode volume simulations for high-$Q$ 2D photonic-crystal cavities. It does not discuss the imaginary part of $\tilde{V}$. Consistently with [17], we will call the normalization in [4] the Leung-Kristensen (LK) normalization, hereafter.

One year after, Christophe Sauvan and his colleagues proposed a very different method to normalize QNMs, promoting the use of the Lorentz reciprocity theorem (divergence theorem) and complex coordinate transforms [5]. They showed that the integral of $\varepsilon \tilde{\mathbf{E}}^2$ over the entire open space is convergent if one uses a complex coordinate transform that damps the field. This is a mathematical result valid for the continuous Maxwellian operator and infinite open spaces. They demonstrate in the Suppl. Mat. that the integral is independent of the coordinate transform, and thus unambiguously defines the mode volume. Considering that such complex coordinate transforms are routinely performed inside perfectly-matched-layers (PMLs), they also propose a numerical method, known as the PML method nowadays, to practically normalize QNMs at complex frequencies with solvers relying on PMLs and verify in the Suppl. Mat. that the normalization is independent of the PML considered. The complex $\tilde{V}$ is introduced by considering the QNM expansion of the field radiated by a dipole source (Green-tensor expansion), and $\text{Im}(\tilde{V})$ is introduced as the dominant correction term to the classical Purcell formula. The rest of the Letter emphasizes the importance of $\text{Im}(\tilde{V})$, evidencing that non-Hermiticity leads to non-Lorentzian LDOS and even negative LDOS for 3D plasmonic and photonic-crystal cavities when $\text{Im}(\tilde{V})$ and $\text{Re}(\tilde{V})$ have similar values.[2]

Complex mode volumes are becoming increasingly popular and used nowadays, especially for plasmonic resonators, which was unthinkable a few years ago. As said in the introduction, the following examines the exactness of the 'extension statement' presenting the work performed in [5] as an extension of [4]. Hopefully, this examination will lead us to discuss some interesting issues related to mode normalization that starts to be consensually accepted nowadays.

---

[2] By LDOS, we mean the local density of electromagnetic states. Note that the QNM decomposition of the LDOS has been extended to the CDOS, by generalizing the definition of the mode volume to enable the study of the spatial coherence in dissipative and resonant photonic systems, see Phys. Rev. A **89**, 043825 (2014).

## 3. Mode normalization: Square integrable QNM fields

As mentioned before, mode volume and mode normalization are intimately linked. Thus, we now consider the literature on QNM normalization, which are not considering complex $\tilde{V}$'s issues but are important to assess the exactness of the 'extension statement' in Section 5. A careful screening of the literature on the normalization issue is beyond the scope of this report. We will however consider (expectedly) all relevant information related to the 'extension statement'.

The question of the spectral reconstruction of waves as a superposition of QNMs, and therefore of the normalization of QNMs, has a venerable history, which probably started in quantum mechanics in the first half of the last century. A review of the different methods used to force the QNM fields to be square integrable by carrying out a suitable mathematical transformation can be found in the Chapter 5 in [18]. Typically, these methods somewhat amount to implement an appropriate rotation of the coordinate (a complex coordinate transformation) along which the divergence occurs into the complex frequency plane. Famous reconstruction examples in quantum mechanics are the 1D tunnel junction and the scattering of an electron in a spherical potential, two systems for which the QNM field is *analytically* known in all space. The complex coordinate transforms have been also applied in electromagnetism, initially for the electromagnetic analogues of the quantum examples, the 1D Fabry-Perot resonator and the scattering by a Mie sphere, see for instance [19,16]. These analytical cases will not be further discussed. The question is how to properly define the volume integral of Eq. (2b) when the QNM fields are not known analytically, for instance for a photonic crystal cavity, a nanoantenna …

In the 90's, the Hong Kong group played a pioneering role in the development of QNM electromagnetic theory. They derived important theorem on the completeness of QNM expansions initially for non-dispersive systems and then for dispersive ones. They studied first [20] and second [16] order QNMs perturbation theory for shape or permittivity changes, a work that consists in finding the new QNMs of a perturbed system by expanding their fields in the basis of the QNMs of an initial unperturbed system. Although their works were indeed requiring QNMs normalization, Leung and his coworkers seem to have rarely discussed $\tilde{V}$ issues *directly*, except in [21] in which the authors present consequences of their work for the second quantization theory. Clearly, the Hong Kong group has correctly treated the case of spheres, but the work lacks from generality. We will come back on generality issues later on.

The PML normalization in [5] is an incarnation of the generalized version in 3D and in electromagnetism with vector fields of the complex coordinate transform introduced in quantum mechanics for simple systems and scalar fields. Nowadays, the term 'PML' is often used to refer to a numerical technique (or trick) that mimics a complex coordinate transform following the equivalence between material parameters and spatial coordinate transforms [22,23]. Primarily, a PML is a mathematical complex coordinate transform [24] that maps a new finite space onto the initial semi-infinite open space, while satisfying the outgoing wave conditions for propagative and evanescent fields. In their simplest forms, numerical PMLs consist in a series of anisotropic, dispersive (or not) and magnetic layers that are often optimized for the sake of numerical accuracy.

I again emphasize that the PML normalization not only provides a numerical method to normalize QNMs, simply by performing the volume integral of Eq. (2b) in the numerical PML layer, but also provides mathematical mappings satisfying the outgoing wave conditions for many 3D geometries of general interest [25]. An example of how the equivalence may be used and even combined can be found in [26].

The PML normalization provides an analytical expression for the response of a non-conservative system driven at complex frequencies asymptotically close to the resonance, see [27] and Section 4.2 in [2]. By exploiting this expression, in 2013, Bai and his colleagues developed another method [27] to normalize QNMs based on the frequency-pole expansions (residue theorem) and launched the freeware QNMPole.[3] The method is very general: it can be implemented with any frequency-domain electromagnetic solver (using PMLs or not), with any dispersive materials (as long as an analytical expression of the permittivity is known), and for 'any' geometries including complicated microcavities that leaks into semi-infinite periodic waveguides, a situation successfully tested in [29] and often encountered in photonic-crystal integrated optics. It would be interesting to compare the residue theorem normalization [27] with the approach developed in [30].

Another important actor in the field in the early 2010's is the Cardiff group [31-34,17]. Egor Muljarov, Wolgang Langbein and colleagues who were developing an exact (up to any order theoretically) QNM perturbation theory known as the resonant-state-expansion. For 2D and 3D systems, they were also using the divergence theorem to derive a normalization formula based on surface and volume integrals, see Eqs. (3) and (4) in [33] or Eq. (4) and Appendix A in [34]. Both the surface and the volume integrals diverge as the integration volume tends to infinity, but their superposition, however, removes the divergencies, making the normalization independent of integration volume. The surface integral relies on an analytic continuation close to the resonance, which is calculated by a Taylor expansion in the spherical coordinates under the assumption that the resonator is embedded in a uniform background. Albeit technical, this last remark will have its importance to interpret the numerical results in the next Section. Let us note that the limitation due to background uniformity has been recently removed and that the "new general formula", which is also more complex, coincides with the PML normalization,[4] an important result that lifts recurrent doubts or inconsistencies from the literature from the last 10 years.

Another option consists in considering the computer world of linear algebra problems with huge matrices of *finite* dimensions, after discretization of the continuous operator. Orthonormalisation is then ensured by bi-orthogonal products and the use of right and left eigenvectors, see Chapter 6 in [18]. When the right and left eigenvectors are computed with a numerical technique that uses PMLs, the PML normalization is naturally implemented. Albeit intuitive, this result has received a recent proof in [35]. This approach has been used in [36], to study the Purcell factor of photonic crystal cavities. In my opinion, the numerical tool developed in [36] to expand the Green tensor contains all the ingredients to correctly handle the exponential growth of the QNM field with a normalization based on the field in the PML.[5] With respect to purely numerical issues only, the approach adopted in [5] is similar (the Lorentz reciprocity normalization and the right-left eigenvector normalizations are equivalent [35]). For dispersive media, the eigenproblem is no longer linear and the right-left eigenvector orthonormalization becomes impossible. In [5], it is shown with the Lorentz reciprocity theorem that normalization is still possible but that the lack of orthogonality leads to an absence of analyticity for the modal coefficients of the QNM expansion, a fact numerically evidenced in Fig. 8a in [2]. Convincingly accurate reconstructions of the scattered field with QNM expansions augmented by PML modes[6] were first provided in [37] for the 2D non-dispersive case. In [25], accurate reconstructions have been reported for various 3D dispersive nanoresonators, including the important case of metal substrates. Furthermore, using the Lorentz reciprocity theorem, it was shown that QNM/PML-mode orthogonality is recovered for dispersive materials that can be mapped into a linear eigenproblem with auxiliary fields. Further recent readings that I have not yet digested and linked are [35,38,39] for instance, but we start to go really beyond the scope of the present report.

---

[3] QNMPole is one of the two QNM solvers of the freeware MAN (Modal Analysis of Nanoresonators) [28].

[4] E. Muljarov and T. Weiss have recently proposed a "general formula for the normalization", see Eq. (20) in Opt. Lett. **43**, 1978 (2018). Actually, it is possible to demonstrate that the new formula is mathematically equivalent to the PML normalization [5] (Haitao Liu, private communication).

[5] Astonishingly, the authors have not exploited the non-Hermiticity brought by the Green tensor extension and have even neglected Im($\tilde{V}$) to end up with a classical Lorentzian dependence of the LDOS, see the discussion related to Eq. 18. They did not discuss the non-dispersive case. The authors of [5] were not aware of [36] and unfortunately did not quote this paper.

[6] The term 'PML modes', introduced in [37], encompasses eigenmodes due to the coupling with the open-space reservoir (possibly with branch cuts for substrate cases, see [25] and Opt. Lett. **44**, 3494 (2019) for striking examples), 'numerical' eigenmodes due to the imperfect discretization, without mentioning accumulation points… Therefore, the term PML modes, which is obviously misleading, should be understood in a generic sense, see Fig. 7a in [2]. I emphasize that any numerical method, even integral methods that do not need PMLs, have PML modes in this sense: in numerical practice, the QNMs cannot be computed over the entire electromagnetic spectrum from $]-\infty;+\infty[$ and completeness is guaranteed by including all these 'numerical' modes.

## 4. Mode volume: comparison of different formulas available in 2014

In 2014, my colleagues in Palaiseau and myself were aware of the availability of different normalization formulae. We were wondering about potential differences between them. From the previous Section, it is easy to realize that it was challenging to understand the link between all the normalizations. The LK normalization, the Cardiff normalization and the PML normalization were all relying on similar approaches, the divergence theorem, but still were significatively different. The residue-theorem normalization was completely different since it does not rely on any volume integral.

Therefore, we decided to resort to fully numerical testing of the formula. We took a 2D example to lower numerical inaccuracies. We chose a plasmonic patch antenna geometry because it was increasingly popular at that time and because the exponential growth of the field occurs typically at distances $\sim Q\lambda$ from the resonator, making high-$Q$ cavities less relevant for benchmarking. The results obtained in 2014 are shown in Fig. 1. We started writing an article but decided not to submit it essentially because the numerical results, albeit interesting, were not accompanied by an in-depth interpretation. The work was fortuitously published in a SPIE proceeding [40] on the occasion of a conference given by Christophe Sauvan in San Diego.

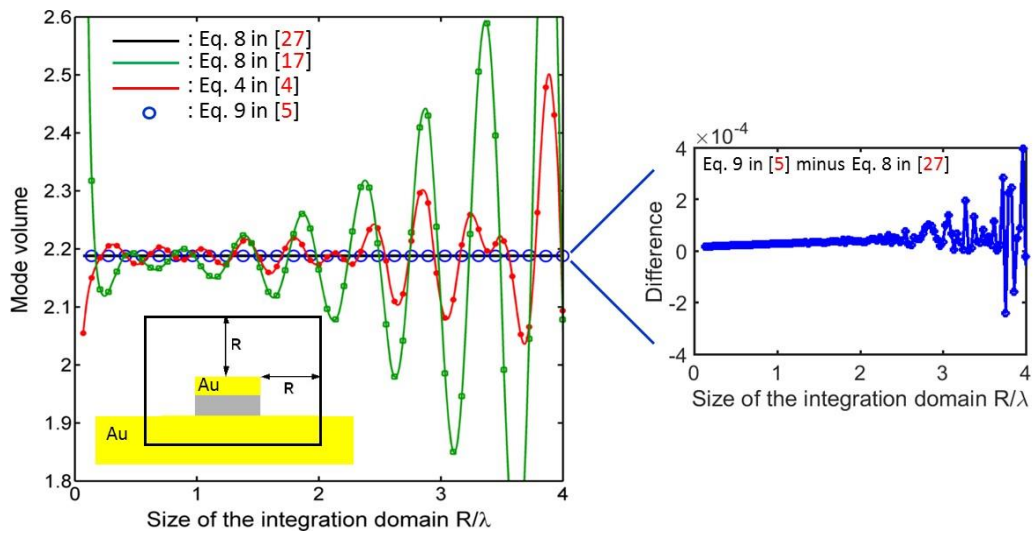

Figure 1. Comparison of several normalization methods available in 2015. The left panel is taken from [40] and the left zoom is obtained with the data in [40]. Only the real part of the mode volume is considered. **Left inset**: 2D nanogap antenna geometry used for the test. The antenna is assumed to be composed of Drude gold metals surrounding a 30-nm thick and 100-nm long dielectric spacer with a refractive index 1.5. **Main panel**: Four normalization methods are tested as the size $R$ of integration domain is varied. The residue-theorem method in [27] does not depend on any integration domain and thus the mode volume does not depend on $R$; it is represented by the horizontal black solid line. The mode volume unit is $\frac{3}{4}(\lambda/10)^2$ with $\lambda = 750$ nm. Equation 8 in [17] was published in 2016 but was available in a preliminary arXiv version in 2014.[7] Further note that the PML and residue-theorem normalizations have not received any refinement or improvement since 2013. They are both available in the freeware MAN (Modal Analysis of Nanoresonators) [28]. **Right inset**: Relative difference between $\text{Re}(\tilde{V})$ computed with the PML and the residue-theorem

---

[7] The integration domain used for all methods is a rectangle box (left inset). We thus used Eq. (8) in [17] which may be limited to uniform background cases (without a substrate for instance), as mentioned later, rather than earlier Cardiff work [34] which requires a spherical integration domain.

normalizations. When assessing the difference values, we need to consider that every point in the inset comes from an independent full-wave computation with different 'discretization' and numerical PMLs. For all computations, the number of Fourier harmonics is 2001, implying that the discretization grain decreases as the size of the computational window $R$ increases. This explains the slight initial slope of the curve for small $R$ followed by stronger chaotic oscillations at large $R$. In our opinion, the variations can be considered as extremely small owing to the fact that $R$ is varied by a factor 30 (almost a factor 1000 for the surface integral used to compute the normalization integral in Eq. 2b).

For the test, special care was devoted to compute the surface integral terms as accurately as possible, especially for the Cardiff formula for which the surface integrand relies on non-trivial gradients normal to the surface. The numerical data have all been obtained with the same numerical method, the aperiodic Fourier Modal Method (a-FMM) [41,26], an in-house method promoted by my group in 2001. In 2014, thirteen years after, we already had a long experience with the method. Note that we also generalized the LK and Cardiff normalizations to dispersive materials. We can provide the proceeding [40] on request.

Albeit incomplete in the sense that only a single example is provided, I do think that the graph in Fig. 1 is an instructive historical testimony of the state-of-the-art of the field in 2014. The comparison evidences many difficulties and subtleties:

- For the Residue-theorem normalization [27], there is no integration volume. So, there is a single value shown with the horizontal line. The value has been obtained with the QNMPole solver [27] of the freeware MAN. The numerical data can thus be reproduced nowadays, just by using the software.

- All the points obtained with the PML normalization are horizontally aligned, consistently with the numerical data reported in the Suppl. Info. in [5]. This is yet another numerical evidence that the normalization obtained with the PML method are independent of the PML used. However, in the present case, the example is related to a complicated geometry with a dispersive substrate.

- All the points obtained with the PML normalization are superimposed with the Residue-theorem normalization value. I emphasize that, though they are obtained with the same numerical method (the a-FMM), the normalization implementations are *completely* different. One method relies on the computation of the volume integral of Eq. (2b) inside the PML, whereas the other method just relies on the computation of the field at a single point in space. The quantitative agreement between two largely different methods is an impressive evidence of consistency and practicality. It gave us great confidence on the soundness of our approaches in 2014.

- The Cardiff and LK normalizations are apparently diverging as the integration domain increases. We also see that they provide *significantly* different results. This is not astonishing as the surface integral terms used to compute the volume integral of Eq. (2b) are different (for a 2D case the surface integrals are linear integrals in fact).

- Nowadays, we know why the Cardiff normalization provides incorrect results for the patch-antenna example. It is because the geometry encompasses a substrate, while the theoretical derivation in [17,33-34] requires asymptotic ($R \to \infty$) forms of the QNMs fields which are only available for uniform backgrounds. It is easy to understand why the presence of a substrate represents a difficulty for establishing a theoretical method to normalize QNMs. To force the QNM fields to be square integrable, it is crucial to dispose of an analytical form of the field, or at least an asymptotic form for $r \to \infty$. This is not possible with a substate, even for a simple air/glass interface or this would demand considerable analytical efforts.

- Why is the LK normalization diverging? The presence of a substrate is one reason, but this is not the only one. The divergence would exist even when the nano patch is in a uniform background. This is discussed in the next Section.

To my knowledge, Fig. 1 represent the first normalization published for a resonator mode in the presence of a substrate (additionally dispersive). The second is found in [42], with a study of a similar 3D patch antenna to clarify experimental results reporting Purcell factors of 1000 with a silver nanocube antenna on a gold substrate [43]. Much more recently, the same 3D patch geometry was used to benchmark several numerical methods to compute and normalize QNMs, and an excellent agreement was achieved by many different contributors, numerical methods, meshes and PML layers.[8]

## 5. Discussion

In 2014, we were not the only ones interested in comparing the different normalizations. A year after, Kristensen and coworkers published a comparison of three methods, the LK, PML and Cardiff normalizations, to conclude that (first sentence of the abstract in [44])

> *«We discuss three formally different formulas for normalization of quasinormal modes currently in use for modeling optical cavities and plasmonic resonators and show that they are complementary and provide the same result. ».*

One easily imagines that, in view of the results of Fig. 1 that were available in my group one year before, I was really surprised when reading this conclusion. I should add that the article contains *wrong* statements on the mathematical equivalence between the LK and PML normalizations. These statements were already presented in the Appendix C of an article published by the same authors in New Journal of Physics the year before. Aside any numerical considerations that may provide the impression that different normalizations provide the same results, this equivalence, see the last paragraph at the end of Section 2B in [44], is not proven at all. The equivalence is based on the false argument that since the PML normalization integral is independent of the coordinate transform as shown in [5], the normalization «*integral must have a well-defined value also under the trivial transformation where no coordinate rotation is performed*». This is a nonsense: the PML normalization is by essence correct when the damping coefficient of the coordinate transform is strong enough and quenches the exponential growth of the QNM field.

Complementary in-depth discussions on the divergence in the absence of PML damping are further found in [45].

The conclusions in [44] have been vividly debated in the literature in a series of articles, comment and reply [45,46]. The publication [45] has been considerably delayed by the Editor of Phys. Rev. A, and thus it is also instructive to look at the arXiv various versions of [45,46]. In their comment, in relation with the conclusion in [44], Egor Muljarov and Wolfgang Langbein state (second sentence of the abstract in [45])[9]

> *« We show here that this conclusion is incorrect and illustrate that the normalization of P. T. Kristensen et al., Opt. Lett. **37**, 1649 (2012), is divergent for any optical mode having a finite quality factor ».*

---

[8] The benchmark, see J. Opt. Soc. Am. A **36**, 686-704 (2019), not only contributes to the elaboration of standards for the normalization of resonances, but also evidences that the divergence of QNM fields is under control nowadays, despite some publications that too quickly suggest that the spatial QNM-field divergence is either unphysical or a big issue for the convergence of QNM expansions. Note that since QNMs are exponentially damped in time, they exponentially grow as $t \to -\infty$. They also exponentially grow in space as $r \to \infty$, but this divergence is no more unphysical than the divergence in time. The growth is observed only over finite intervals in time or in space owing to causality. Interesting discussions can be found in OSA Continuum **1**, 340 (2018) and Phys. Rev. B **98**, 085418 (2018).

[9] The conclusion in [45,17] puts serious doubts on the generality of the pioneer work of the Hong Kong group. The latter was mainly considering 1D and spherical resonators. The 1D case does not present any difficulty for the normalization, and for some reasons that I have not precisely grasped (probably in relation with the orthogonality of spherical harmonics), it appears that the LK normalization is only correct for spherical resonators in uniform media, for which numerous numerical evidences are reported in Leung's papers in the 90's.

A gentle summary of the vivid debate is found in Section 4.1 in [2]. With less tactful and more sincere wordings now, I would say that the articles [44,46] try to "noyer le poisson" (a French expression that seems to mean "evade or cloud the issue"). Indeed, a numerical example is not a proof. However, a single numerical example evidencing a malfunction by a counterexample, like in Fig. 1, is sufficient to prove that a theoretical result is incorrect. The results of Fig. 1 clearly evidence that none of the LK, PML or Cardiff normalizations «*provides the same result*». They are not «*complementary*» neither. In experiments, there is always a subtle difference between what has been really observed and what is said to be demonstrated. In theory, it is clearer. In the present case, the normalization and the mode volume proposed in [4] are both incorrect, and factually [4] results in nothing more than replacing the Hermitian formula, which is not intellectually satisfactory albeit successfully used plenty of times for high-$Q$ systems, by another formula that is not more intellectually satisfactory and is further diverging, completely incorrect to predict $\text{Im}(\tilde{V})$, and probably does not bring much concerning $\text{Re}(\tilde{V})$ compared to the Hermitian formula.

## 6. Epilogue

Admittedly, the 2012 6-formula Letter [4] is much easier to read than the 2013 PRL Letter [5]. For novices who start in the field with the Tutorial [3] and are not aware of normalization subtilities, it is likely that they retain that [5] is simply an extension of earlier works. It is an uninstructive shortcut that masks the complexity of the normalization issue. This is also misleading as it carries towards wrong directions with major flaws which will be inevitably echoed by others. An echo example is found in a recent review on biosensing with plasmonic nanoresonators [47], where the issue related to a correct non-Hermitian definition of the mode volume is discussed in the beginning of the review (see Eqs. 1 and 2 therein) by recommending further readings of two Kristensen's articles [4] and [9] that contain obvious flaws on the mode volume definition. It is unfortunate since articles [15,48] or reviews [2] properly address the normalization issue exactly in the context of sensing with plasmonic nanoresonators.

When we submitted the review article [2] in 2017, we also posted an arXiv and invited the key actors in the field to give their first impressions. We also did not suggest to the editor of Laser & Photonics Reviews to exclude any reviewer and received controversial comments that helped improving the manuscript. Simply we though that a review could be commented by all before publication and should try to present a consensual view of the field. On another hand, none of the authors of the review article [2] reviewed the Tutorial (personal communications). Egor Muljarov, another important actor in the field, neither (personal communication). Indeed, the Tutorial was available on arXiv long before its publication. I have not been invited to comment on it. However, I have contacted the last author[10] at least twice in December 2019 to engage a discussion on several points of concern. The emails have not been answered.

Nowadays, people are interested in computing and normalizing QNMs for various geometries. Presenting [5] as an extension of [4] in 2020 is a nonsense in my opinion, or at best a misleading stylistic exercise.

There are many interesting situations requiring a proper definition of the mode volume to help interpreting experiments [49,50,51]. There are also interesting basic works aiming at understanding the complex nature of $\tilde{V}$'s nowadays. Both the imaginary and real parts of $\tilde{V}$ have been recently measured for a photonic-crystal cavity [52]. The $1/\tilde{\mathbf{E}}^2$ dependence that encompasses twice the phase of the mode at the position of the perturber has been physically explained [52] and the impact of the imaginary part of $\tilde{V}$ on the negative contribution of some QNMs to the LDOS has been verified experimentally [53]. Preliminary discussions on the impact of $Im(\tilde{V})$ on strong coupling between a nanoresonator and a quantum emitter are also available [2]. And above all, there are open questions, for instance on how to extend the definition of the mode volume when quantum surface correction cannot be neglected [54] or when localized fields appreciably vary at the scale of the molecular orbitals [55]. Further interesting works are ahead.

---

[10] For obvious reasons, I thought it was preferable not to contact Philip Kristensen directly and preferred to contact Kurt Busch. He is about my age and I know him personally.

**Acknowledgements**

I would like to acknowledge Christophe Sauvan and Jean Paul Hugonin for their multiple contributions to the work and analysis presented in Fig. 1. Christophe Sauvan has prepared the right inset of Fig. 1 with the data available in 2015. I acknowledge helpful comments of Egor Muljarov, especially concerning subtle differences between the KL and Cardiff normalizations and clarifications of the applicability of Cardiff normalization. I am grateful to Haitao Liu for relevant suggestions and for pointing to my attention that the "general formula for the normalization" recently published by Egor Muljarov and Thomas Weiss is mathematically equivalent to the PML normalization, see note 4.